\documentclass[12pt]{article}

\begin{document}

\title{\bf Conformal Ricci and Matter Collineations for Anisotropic Fluid}

\author{M. Sharif \thanks{msharif@math.pu.edu.pk} and Naghmana Tehseen\\
Department of Mathematics, University of the Punjab,\\
Quaid-e-Azam Campus, Lahore-54590, Pakistan.}
\date{}

\maketitle
\begin{abstract}
We study the consequences of timelike and spaccelike conformal Ricci
and conformal matter collineations for anisotropic fluid in the
context of General Relativity. Necessary and sufficient conditions
are derived for a spacetime with anisotropic fluid to admit
conformal Ricci and conformal matter collineations parallel to $u^a$
and $x^a$. These conditions for timelike and spacelike conformal
Ricci and conformal matter collineations for anisotropic fluid
reduce to the conditions of perfect fluid when the heat flux and the
traceless anisotropic stress tensor vanish. Further, for $\alpha=0$
(the conformal factor), we recover the earlier results of Ricci
collineations and matter collineations in each case of timelike and
spacelike conformal Ricci collineations and conformal matter
collineations for the perfect fluid. Thus our results give the
generalization of the results already available in the literature.
It is worth noticing that the conditions of conformal matter
collineations can be derived from the conditions of conformal Ricci
collineations or vice versa under certain constraints.
\end{abstract}

%\newpage

\section{Introduction}

In General Relativity (GR), symmetries are used to understand the
natural relationship between geometry and matter given by the
Einstein field equations (EFEs). Symmetric system has not only the
advantage of certain simplicity or even beauty but also special
physical effects frequently occur. The symmetries are important in
the classification of spacetime. Symmetries of geometrical/physical
quantities are known as collineations. A symmetry (collineation) is
defined by a relation
\begin{eqnarray*}
\pounds_\xi\phi=A,
\end{eqnarray*}
where $\pounds$ is the Lie derivative operator, $\xi$ is the
symmetry or collineation vector, $\phi$ is any of the quantities
$g_{ab},~R_{bcd}^a,~R_{ab},~\Gamma^a_{bc},~T_{ab}$ and geometric
objects constructed from them, $A$ is a tensor with the same
symmetries as $\phi$. When $\phi_{ab}=g_{ab}$ and $A_{ab}=2\alpha
g_{ab}$, the symmetry vector $\xi$ is called conformal Killing
vector (CKV) and specializes to KV when $\alpha=0$. When
$\phi_{ab}=R_{ab}$ and $A_{ab}=\alpha R_{ab}$, the symmetry vector
$\xi$ is called conformal Ricci collineation (CRC) or Ricci
inheritance collineation and specializes to Ricci collineation (RC)
for $\alpha=0$.
\begin{equation}
\pounds_{\xi} R_{ab} =\alpha R_{ab}.
\end{equation}
When $\phi_{ab}=T_{ab}$ and $A_{ab}=\alpha T_{ab}$, where $T_{ab}$
is the energy momentum tensor, the vector $\xi$ is called
conformal matter collineation (CMC) or matter inheritance
collineation and becomes matter collineation (MC) when $\alpha=0$.
\begin{equation}
\pounds_{\xi} T_{ab} =\alpha T_{ab}.
\end{equation}
The function $\alpha$ in the case of CKVs is called conformal factor
and in the case of conformal or inheriting collineations the
conformal or inheriting factor. It is mentioned here that we are
using the term \emph{conformal Ricci or conformal matter
collineation} instead of \emph{Ricci or matter inheritance
collineation}.

The study of inheritance symmetries with CRC and CMC in fluid
spacetimes has recently attracted some interest. Oliver and Davis,
[1,2] gave necessary and sufficient conditions for a matter
spacetime to admit an RC.  Letelier [3] discussed anisotropic fluids
with two perfect fluid component. Herrera and Ponce [4-6] studied
CKVs with particular reference to perfect and anisotropic fluids
with different combination of fluids. Maartens et al. [7] made a
study of SCKVs in anisotropic fluid. Carot et al. [8] investigated
spacetime with CKVs. Coley and Tupper [9] studied spacetimes
admitting SCKV and symmetry inheritance. Duggal [10,11] discussed
curvature inheritance symmetry and timelike CRC in perfect fluid
spacetime. The theory of spacelike congruence in GR was first
formulated by Greenberg [12] who discovered its application to the
vortex congruences in a rotational fluid. This was developed further
and applications to spacelike CKVs in spacelike congreuence were
considered by Mason and Tsamparlis [13]. Yavuz and Yilmaz [14] and
Yilma et al. [15] considered CKVs and SCKVs and worked on the
conformal curvature symmetry in string cosmology. Yilmaz [16] also
studied timelike and spacelike collineations in string cloud.

Tsamparlis and Mason [17] investigated RCs in fluid spacetimes
(perfect, imperfect and anisotropic fluids). Tsamparlis [18]
discussed the conditions on the kinematic quantities of the
congurence of the vector field generating the collineation. Baysal
and Yilmaz [19] worked on spacelike conformal Ricci collineation
(SpCRC) in a model of string fluid and string stress tensor. The
same authors [20] also studied timelike and SpRCs in the model of
string fluid. Sariddakis and Tsamparlis [21] discussed the
applications for SpCKV and matter described either by perfect fluid
or by an anisotropic fluid. Recently, Tsamparlis [22] discussed
general symmetries of string fluid spacetime. In a recent paper,
Sharif and Umber [23] have investigated timelike and SpCMC in
specific forms of the energy-momentum tensor. They discussed the
necessary and sufficient conditions for a vector to be timelike and
spacelike admitting CMC.

The main purpose of such kind of work is the simplification of the
EFEs to find the exact solutions. We describe the geometrization of
a symmetry in terms of necessary and sufficient conditions on the
geometry of the integral lines of the vector field that generates
the symmetry. This enables us to express symmetry in a form which is
convenient to the simplification of the field equations in a direct
and inherent way. The symmetry can be studied in most convenient way
if we take the Lie derivative of the field equations with respect to
the vector $u^a$ or $\xi^a$ that generates the symmetry. This yields
an expression such that the left hand side contains the Lie
derivative of the Ricci tensor and the right hand side has the Lie
derivative of the energy-momentum tensor. Consequently, we have the
field equations as Lie derivatives along the symmetry vector of the
dynamical variables.

In this paper, we shall elaborate the necessary and sufficient
conditions for the existence of timelike and spacelike CRC and CMC
for anisotropic fluid. The layout of the paper is the following. In
section $2$, we review the $1+1+2$ decomposition and consider the
decomposition of the quantities which will be used in later
sections. In section $3$, we investigate the kinematic conditions
for timelike and spacelike CRCs and CMCs. Section $4$ is devoted to
derive the necessary and sufficient conditions for anisotropic fluid
spacetime which admit CRCs. In section $5$, we derive these
conditions for anisotropic fluid spacetime which admit CMCs.
Finally, section $6$ contains summary and discussion of the results.

\section{Notation}

Let us consider a spacelike four position vector $x^a$ and the
corresponding unit timelike four-velocity vector $u^a$. Both the
vectors are perpendicular to each other satisfying the relations
given as follows
\begin{equation}{\setcounter{equation}{1}}
u^au_a=-1,\quad u^ax_a=0,\quad x^ax_a=1.
\end{equation}
The kinematical quantities $\sigma_{ab},~\omega_{ab},~\dot{u}_a$ and
$\theta$ are defined [24] as
\begin{eqnarray}
\dot{u}_a &=& u_{a;n}u^n=\frac{Du_a}{D\tau},\\
\omega_{ab} &=& u_{[a;b]}+\dot{u}_{[a}u_{b]},\\
\sigma_{ab} &=& u_{(a;b)}+\dot{u}_{(a}u_{b)}
-\frac{\theta}{3}h_{ab},\\
\theta &=& u^a_{;a}.
\end{eqnarray}
Using these quantities, the covariant differentiation of the
four-vector velocity can be written in $1+3$ decomposition as
follows
\begin{equation}
u_{a;b}= \omega_{ab}+\sigma_{ab}
+\frac{\theta}{3}h_{ab}-\dot{u}_au_b.
\end{equation}
The projection tensor, $h_{ab}=g_{ab}+u_au_b,$ has the following
properties:
\begin{equation}
h^{cd}x_c=x^d,\quad h^{cd}u_c=0,\quad h_c^d h_d^b=h_c^b,\quad
h_{cd}=h_{dc},\quad h_c^c=3.
\end{equation}
If we use $\xi^a=\xi u^a$, the conformal Ricci and matter
symmetries, respectively, can be re-written as
\begin{eqnarray}
\dot{R}_{ab} + 2u^cR_{c(a}ln\xi_{,b)}+2R_{c(a} u^c_{;b)} = \beta
R_{ab},\\
\dot{T}_{ab} +2u^cT_{c(a} ln\xi_{,b)} + 2T_{c(a} u^c_{;b)} = \beta
T_{ab},
\end{eqnarray}
where $\beta=\frac{\alpha}{\xi}$. We define the screen projection
operator normal to both $u^a$ and $~x^a$ as
$$H_{ab}=h_{ab}-x_ax_b$$ which obeys the following
properties:
$$H_{ab}u^a=0=H_{ab}x^a,\quad H_{ab}h_{c}^b=H_{ac},\quad
H^a_{a}=2.$$ The $1+1+2$ decomposition of $x_{a;b}$ can be given as
follows [25]
\begin{equation}
x_{a;b}=A_{ab}+x^*_ax_b-\dot{x}_au_b+u_a[x^tu_{t;b}
+(x^f\dot{u}_f)u_b-(x^fu^*_f)x_b],
\end{equation}
where $s^*=s_{...;a}x^a$ and $A_{ab}=H_a^cH^d_bx_{c;d}.$ The
decomposition of $A_{ab}$ into its irreducible parts is
\begin{equation}
A_{ab}=S_{ab}+B_{ab}+\frac{1}{2}\varepsilon H_{ab},
\end{equation}
where $S_{ab}=S_{ba},~S_b^b=0$ is the traceless part (shear tensor),
$B_{ab}=-B_{ab}$ is the antisymmetric part (rotation tensor) and
$\varepsilon$ is the trace (expansion). These quantities are given
as
\begin{equation}
S_{ab}=(H^c_aH^d_b-\frac{1}{2}H^{cd}H_{ab})x_{(c;d)},\quad
B_{ab}=H^c_aH^d_bx_{[c;d]},\quad \varepsilon=H^{cd}x_{c;d}.
\end{equation}
We can write the square bracket term in Eq.(2.10) as
\begin{equation}
-N_{b}+2w_{fb}x^f+H^f_{b}\dot{x}_f,
\end{equation}
where $N_a$ is given by
\begin{equation}
N_a=H^b_a(\dot{x}_b-u^*_b).
\end{equation}
This is called Greenberg vector [25]. Using Eq.(2.13) in Eq.(2.10),
it follows that
\begin{equation}
x_{a;b}=A_{ab}+x^*_ax_b-\dot{x}_au_b+H_{b}^c\dot{x_c}u_a
+(2w_{tb}x^t-N_{b})u_a.
\end{equation}
Also, we have
\begin{equation}
x^fu_{f;b}=2x^fu_{[f;b]}+u^*_b
=-2\omega_{bf}x^f-(x_f\dot{u}^f)u_b+u^*_b.
\end{equation}
When $\xi^a=\xi x^a$, the conformal Ricci and conformal matter
symmetries can be written as
\begin{eqnarray}
R^*_{ab}+2x^cR_{c(a}\ln\xi_{,b)} +2R_{c(a}x^c_{;b)}=\beta R_{ab},\\
T^*_{ab}+2x^cT_{c(a}\ln\xi_{,b)} +2T_{c(a}x^c_{;b)}=\beta T_{ab}.
\end{eqnarray}
where $\beta=\frac{\alpha}{\xi}$.

The energy-momentum tensor of anisotropic fluid is given by [24]
\begin{equation}
T_{ab}=\rho u_au_b+ph_{ab}+2q_{(a}u_{b)}+\pi_{ab},
\end{equation}
where $\rho$ is the total energy density, $p$ denotes the isotropic
pressure, $q_a$ is the heat flux vector and $\pi_{ab}$ is the
traceless anisotropic stress tensor. The quantities $u_a, q_a$ and
$\pi_{ab}$ satisfy the following relations:
\begin{eqnarray*}
u^aq_a=0,\quad \pi_{ab}u^a=0,\quad \pi^a_a=0.
\end{eqnarray*}
Using the Einstein field equations
\begin{eqnarray*}
R_{ab}-\frac{1}{2}Rg_{ab}+\Lambda g_{ab}=T_{ab},
\end{eqnarray*}
where $\Lambda$ is a cosmological constant, the Ricci tensor of
anisotropic fluid can be written as
\begin{equation}
R_{ab}=(\rho+p)u_au_b++\frac{1}{2}(\rho-p+2\Lambda)g_{ab}+2q_{(a}u_{b)}+\pi_{ab}.
\end{equation}

\section{Timelike and Spacelike Conformal Ricci \\Collineations}

This section is devoted to prove the necessary and sufficient
conditions for the existence of timelike and spacelike CRC in the
model of anisotropic fluid. In addition, we shall give the
conditions for the existence of timelike and spacelike CRC in the
form of kinematical quantities.\\
\par\noindent
\textbf{Proposition 1:}
\par\noindent
Anisotropic fluid spacetime with Ricci tensor, given by Eq.(2.20),
admits a timelike CRC $\xi^a=\xi u^a$ if and only if
\begin{eqnarray}{\setcounter{equation}{1}}
\dot{\rho}&+&3\dot{p}+2(\rho+3p-2\Lambda)(\ln \xi \dot{)}
=\beta(\rho+3p-2\Lambda),\\
(\rho &+& 3p-2\Lambda)[\dot{u}_a-(\ln\xi)_{,a}-(\ln\xi\dot{)}u_a]
+2\dot{q}_a+2q_a(\ln\xi\dot{)}\nonumber\\
&+&2\dot{q}_bu^bu_a+2q^b\omega_{ab}+2q^b\sigma_{ab}+\frac{2}{3}\theta
q_a=2\beta q_a,\\
\dot{\rho}&-& \dot{p}+\frac{2}{3}(\rho-p
+2\Lambda)\theta+\frac{4}{3}\dot{u}_fq^f
-\frac{4}{3}q^f(\ln\xi){,f}
+\frac{4}{3}\pi^{bf}\sigma_{fb}\nonumber\\&=&\beta(\rho-p+2\Lambda),\\
(\rho&-&p+2\Lambda)\sigma_{ab}+2q_{(a}\dot{u}_{b)}
-\frac{2}{3}q^f\dot{u}_f
h_{ab}-2q_{(a}(\ln\xi)_{,b)}-2u_{(a}q_{b)}(\ln\xi\dot{)}\nonumber\\
&+&\frac{2}{3}\theta
\pi_{ab}+\frac{2}{3}h_{ab}q^f(\ln\xi)_{,f}+\dot{\pi}_{ab}
+2\dot{\pi}_{f(a}u_{b)}u^f+2\pi_{f(a}\omega^f_{b)}+2\pi_{f(a}\sigma^f_{b)}\nonumber\\
&+&2\pi_{f(a}\sigma^f_{b)}-\frac{2}{3}h_{ab}\pi^{cd}\sigma_{cd}=\beta
\pi_{ab}.
\end{eqnarray}
\par\noindent
\textbf{Proof:}
\par\noindent
First we assume that timelike CRC exists in this spacetime and prove
that the conditions given by Eqs.(3.1)-(3.4) are satisfied.

When we substitute the value of Ricci tensor for anisotropic fluid
from Eq.(2.20) in Eq.(2.8), we have
\begin{eqnarray}
(\rho&+&p\dot{)}u_a u_b+2(\rho+p) u_{(a}
\dot{u}_{b)}+\frac{1}{2}(\rho-p\dot{)}g_{ab} +2q_{(a}\dot{u}_{b)}
+2\dot{q}_{(a}u_{b)}\nonumber\\
&+&\dot{\pi}_{ab}
-(\rho+3p-2\Lambda)u_{(a}(ln\xi)_{,b)} -2q_{(a}(\ln\xi)_{,b)}
+(\rho-p+2\Lambda)u_{(a;b)}\nonumber\\&+&2q_fu_{(a}u^f_{;b)}
+2\pi_{f(a}u^f_{;b)}=\beta[(\rho+p)
u_au_b+\frac{1}{2}(\rho-p+2\Lambda)g_{ab}\nonumber\\
&+&2q_{(a}u_{b)}+\pi_{ab}].\nonumber
\end{eqnarray}
Using $1+3$ decomposition of $u_{a;b}$, the above expression can be
written as
\begin{eqnarray}
[\frac{1}{2}(\dot{\rho}&+&3\dot{p})
+(\rho+3p-2\Lambda)(\ln\xi\dot{)}]u_au_b
+[(\rho+3p-2\Lambda)(\dot{u}_c
-(\ln\xi)_{,c})\nonumber\\
&+&2(\ln\xi\dot{)}q_c+2\dot{q}_c
+2q^f(\sigma_{fc}+\omega_{fc}+\frac{1}{3}\theta
h_{fc})]u_{(a}h^c_{b)}+\{\frac{1}{2}(\rho-p\dot{)}h_{cd}\nonumber\\
&+&(\rho-p+2\Lambda)(\sigma_{cd}+\frac{1}{3}\theta
h_{cd})+\dot{\pi}_{cd} +\frac{2}{3}\theta \pi_{cd}+2q_{(c}
[\dot{u}_{d)}-(\ln \xi)_{,d)}]\nonumber\\&+&
2\pi_{(fc}[\omega^f_{d)}+\sigma^f_{d)}]\}h^c_ah^d_b=
\beta[\frac{1}{2}(\rho+3p-2\Lambda)u_a u_b
+\frac{1}{2}(\rho-p+2\Lambda)h_{ab}\nonumber\\
&+&2q_{(a}u_{b)}+\pi_{ab}].
\end{eqnarray}
Contracting Eq.(3.5) in turn with $u^a u^b,~u^a h_c^b,~h^{ab}$ and
$h^a_ch^b_d -\frac{1}{3} h^{ab}h_{cd}$, we obtain
\begin{eqnarray}
\dot{\rho}&+&3\dot{p}+2(\rho+3p-2\Lambda)(\ln \xi \dot{)}
=\beta(\rho+3p-2\Lambda),\\
(\rho&+&3p-2\Lambda) h^b_c[\dot{u}_b -(\ln\xi)_{,b}]
+2\dot{q}_c+2\dot{q}_fu^fu_c\nonumber\\
&+&2(\ln\xi\dot{)}q_c+2q^f\omega_{fc}+2q^f\sigma_{fc}
+\frac{2}{3}q_c\theta=\beta q_c,\\
3\dot{\rho}&-&3\dot{p}+2(\rho-p +2\Lambda)\theta+4\dot{u}_fq^f
-4q^f(\ln\xi){,f}\nonumber\\
&+&4\pi^{bf}\sigma_{fb}=3\beta(\rho-p+2\Lambda),\\
(h^a_ch^b_d&-&\frac{1}{3}h^{ab}h_{cd}) [(\rho-p+2\Lambda)\sigma_{ab}
+2q_{(a}\dot{u}_{b)}+\dot{\pi}_{ab}\nonumber\\
&-&2q_{(a}(\ln\xi)_{,b)}+2\pi_{f(a}\omega^f_{b)}
+2\pi_{f(a}\sigma^f_{b)}+\frac{2}{3}\theta \pi_{ab}]\nonumber\\
&=&\beta[h^a_ch^b_d-\frac{1}{3}h^{ab}h_{cd}]\pi_{ab}.
\end{eqnarray}
Since
\begin{eqnarray*}
(h^a_ch^b_d-\frac{1}{3}h^{ab}h_{cd})\sigma_{ab}&=&\sigma_{cd},\\
(h^a_ch^b_d-\frac{1}{3}h^{ab}h_{cd})(q_a\dot{u}_b+q_b\dot{u}_a)
&=&2q_{(c}\dot{u}_{d)}-\frac{2}{3}q^b\dot{u}_bh_{cd},\\
(h^a_ch^b_d-\frac{1}{3}h^{ab}h_{cd})\dot{\pi}_{ab}
&=&\dot{\pi}_{cd}+2\dot{\pi}_{b(c}u_{d)}u^b,\\
(h^a_ch^b_d-\frac{1}{3}h^{ab}h_{cd})
(\pi_{fa}\omega^f_{b}+\pi_{fb}\omega^f_{a})&=&
2\pi_{f(c}\omega^f_{d)},\\
(h^a_ch^b_d-\frac{1}{3}h^{ab}h_{cd})
(\pi_{fa}\sigma^f_{b}+\pi_{fb}\sigma^f_{a})&=&
2\pi_{f(c}\sigma^f_{d)}-\frac{2}{3}h_{cd}
\pi^{ab}\sigma_{ab},\\
(h^a_ch^b_d-\frac{1}{3}h^{ab}h_{cd})
(q_a(\ln\xi)_{,b}+q_b(\ln\xi)_{,a})
&=&2q_{(c}(\ln\xi)_{,d)}+2u_{(c}q_{d)} (\ln\xi\dot{)}\\
&-&\frac{2}{3}h_{cd}q^b(\ln\xi)_{,b},\\
(h^a_ch^b_d-\frac{1}{3}h^{ab}h_{cd})\pi_{ab}&=&\pi_{cd}.
\end{eqnarray*}
Substituting these values in Eq.(3.9), it follows that
\begin{eqnarray}
(\rho&-&p+2\Lambda)\sigma_{cd}+2q_{(c}\dot{u}_{d)}-\frac{2}{3}q^b
\dot{u}_b h_{cd}+\dot{\pi}_{cd}+2\dot{\pi}_{a(c}u_{d)}u^a\nonumber\\
&-&2q_{(c}(\ln\xi)_{,d)}-2u_{(c}q_{d)}(\ln\xi\dot{)}
+\frac{2}{3}h_{cd}q^b(\ln\xi)_{,b}
+2\pi_{f(c}\omega^f_{d)}\nonumber\\
&+&2\pi_{f(c}\sigma^f_{d)}-\frac{2}{3}h_{cd}\pi^{ab}\sigma_{ab}
+\frac{2}{3}\theta\pi_{cd}=\beta\pi_{cd}.
\end{eqnarray}
\begin{description}
\item{$(i)$} Eq.(3.6) is the same as Eq.(3.1).
\item{$(ii)$} Eq.(3.2) is obtained by expanding $h^b_c$ in Eq.(3.7).
\item{$(iii)$} Eq.(3.8) is the same as Eq.(3.3).
\item{$(iv)$} Eq.(3.10) gives the condition (3.4).
\end{description}
Now we shall show that if the conditions (3.1)-(3.4) are satisfied,
then there exists a timelike CRC in anisotropic fluid spacetime. For
this purpose, we consider the left hand side of Eq.(3.5), i.e.,
\begin{eqnarray}
[\frac{1}{2}(\dot{\rho}&+&3\dot{p})+(\rho+3p-2\Lambda)(\ln\xi\dot{)}]u_au_b
+[(\rho+3p-2\Lambda)(\dot{u}_c-\nonumber\\
(\ln\xi)_{,c})&+&2(\ln\xi\dot{)}q_c+2\dot{q}_c
+2q^f(\sigma_{fc}+\omega_{fc}+\frac{1}{3}\theta
h_{fc})]u_{(a}h^c_{b)}+\nonumber\\
\frac{1}{2}(\rho&-&p\dot{)}h_{cd}+
(\rho-p+2\Lambda)(\sigma_{cd}+\frac{1}{3}\theta
h_{cd})+\dot{\pi}_{cd}
+\frac{2}{3}\theta \pi_{cd}\nonumber\\
&+&2q_{(c}[\dot{u}_{d)}-(\ln\xi)_{,d)}] +2\pi_{(fc}[\omega^f_{d)}
+\sigma^f_{d)}]h^c_ah^d_b.\nonumber
\end{eqnarray}
Now we make use of Eqs.(3.1)-(3.3) in the above expression so that
\begin{eqnarray}
(\rho&-&p+2\Lambda)\sigma_{ab}+2q_{(a}\dot{u}_{b)}-\frac{2}{3}q^f
\dot{u}_f h_{ab}+\dot{\pi}_{ab}
+2\dot{\pi}_{f(a}u_{b)}u^f\nonumber\\
&-&2q_{(a}(\ln\xi)_{,b)}-2u_{(a}q_{b)}(\ln\xi\dot{)}
+\frac{2}{3}h_{ab}q^f(\ln\xi)_{,f}+2\pi_{f(a}\omega^f_{b)}\nonumber\\
&+&2\pi_{f(a}\sigma^f_{b)}-\frac{2}{3}h_{ab}\pi^{cd}\sigma_{cd}+\frac{2}{3}\theta
\pi_{ab}+\beta[(\rho+p)u_au_b\nonumber\\
&+&\frac{1}{2}(\rho-p+2\Lambda)g_{ab}+2q_{(a}u_{b)}].
\end{eqnarray}
Using Eq.(3.4), we finally obtain
$$\beta[\frac{1}{2}(\rho+3p-2\Lambda)u_a u_b
+\frac{1}{2}(\rho-p+2\Lambda)h_{ab}+2q_{(a}u_{b)}+\pi_{ab}].$$ This
is equal to the right hand side of Eq.(3.5). Hence the conditions
(3.1)-(3.4) are necessary and sufficient for anisotropic fluid
spacetime to admit a timelike CRC.\\
\par\noindent
\textbf{Proposition 2:}
\par\noindent
Anisotropic fluid with Ricci tensor, given by Eq.(2.20), admits a
SpCRC $\xi^a=\xi x^a$ if and only if
\begin{eqnarray}
(\rho
&+&3p)^*+2(\rho+3p-2\Lambda)x_a\dot{u}^a-4q_ax^a(\ln\xi\dot{)}
-4q^aN_a\nonumber\\
&+&4q_ax^au^*_dx^d=\beta(\rho+3p-2\Lambda),\\
(\rho&-&p+2\Lambda)[x^*_a +(\ln\xi)_{,a}-(\ln\xi)^*x_a]
+2(\pi_{ab}x^b)^* -2\pi^*_{cd}x^cx^dx_a\nonumber\\
&+&2\pi_{cd}x^cx^d(\ln\xi)_{,a}-4\pi_{cd}x^cx^d(\ln\xi)^*
+2\pi_{ab}x^b(\ln\xi)^*+2x^bq_b(\ln\xi)^*u_a\nonumber\\
&-&2\pi_{cd}x^{*c}x^dx_a-2\pi_{cd}\dot{x}^cx^du_a
+2x_c\dot{u}^cq_dx^du_a+2\pi^{cd}x_dA_{ac}
+2q^cx^d\omega_{ad}x_c\nonumber\\&=&2\beta[\pi_{ab}x^b-\pi_{cd}x^cx^dx_a
+q_bx^bu_a],\\
(\rho&-&p+2\Lambda)N_a+2(\rho+3p-2\Lambda)\omega_{ab}x^b
+2q_a\dot{x}^bu_b-2q_cx^c\dot{x}^du_dx_a\nonumber\\
&-&2q^*_a+2q^*_cx^cx_a-2q^*_cu^cu_a
+2\pi^*_{ab}u^b+2\pi_{cd}u^{*c}x^dx_a
-2q_bx^b(\ln\xi)_{,a}\nonumber\\
&-&2q_bx^b(\ln\xi\dot{)}u_a-2q_bx^b(\ln\xi)^*x_a
+2\pi_{ba}x^b(\ln\xi\dot{)}-2\pi_{cd}x^cx^d(\ln\xi\dot{)}x_a\nonumber\\
&+&2\pi_{ab}\dot{x}^b-2\pi_{cd}x^c\dot{x}^dx_a-2q^dA_{ad}
=\beta(-q_a+q_bx^bx_a),\\
(\rho&-&p+2\Lambda)[2(ln\xi)^*-\varepsilon]
+3\pi^*_{cd}x^cx^d+6\pi_{ab}x^bx^a(\ln\xi)^*
+2\pi_{ab}x^{*b}x^a\nonumber\\&-&2\pi^{ab}x_a(\ln\xi)_{,b}
+2x^aq^b\omega_{ab}-2\pi^{ab}A_{ab}=3\beta \pi_{ab}x^ax^b,\\
(\rho&-&p+2\Lambda)S_{ab}+2\pi_{c(a}A^c_{b)}-2\pi^{cd}x_dA_{c(a}x_{b)}
+2\dot{x}^c\pi_{c(a}u_{b)}-2H_{ab}\pi^{cd}A_{cd}\nonumber\\
&+&\pi^*_{ab}+2u^c\pi^*_{c(a}u_{b)}
-2x^c\pi^*_{c(a}x_{b)}+\pi^*_{cd}x^cx^dx_ax_b
-2\pi^*_{cd}x^cu^dx_{(a}u_{b)}\nonumber\\
&+&\frac{1}{2}H_{ab}\pi^*_{cd}x^cx^d
-4x^cq_{(a}\omega_{b)c}+4x^cq_dx^dx_{(a}\omega_{b)c}
+2H_{ab}\omega_{cd}q^cx^d\nonumber\\&+&2x^c \pi_{c(a}(\ln\xi)_{,b)}
-2x^cx^d\pi_{cd}(\ln\xi)_{,(a}x_{b)}+2(\ln\xi\dot{)}\pi_{c(a}u_{b)}x^c\nonumber\\
&-&2\pi_{cd}x^cx^d(\ln\xi\dot{)}x_{(a}u_{b)}
-2x^c(\ln\xi)^*\pi_{c(a}x_{b)}+2x^c\pi_{cd}x^d
(\ln\xi)^*x_ax_b\nonumber\\
&-&H_{ab}[x_c\pi^{cd}(\ln\xi)_{,d} -\pi_{cd}x^cx^d(\ln\xi)^*]
=\beta[\pi_{ab}-2x^c\pi_{c(a}x_{b)}\nonumber\\&+&\pi_{cd}x^cx^dx_ax_b
+\frac{1}{2}H_{ab}\pi_{cd}x^cx^d].
\end{eqnarray}
This can be proved on the same lines as Proposition $1$.

\section{Timelike and Spacelikle Conformal Matter \\Collineations}

In this section we shall state and prove the necessary and
sufficient conditions for the existence of timelike and spacelike
CMC in the model of anisotropic fluid.
\par\noindent
\textbf{Proposition 3:}
\par\noindent
Anisotropic fluid spacetime with the energy-momentum tensor, given
by Eq.(2.19), admits a timelike CMC $\xi^a=\xi u^a$ if and only if
the following conditions are satisfied.
\begin{eqnarray}{\setcounter{equation}{1}}
\dot{\rho}&+&2\rho(\ln\xi\dot{)}=\beta \rho,\\
\rho[\dot{u}_a&-&((\ln\xi)_{,a}-(\ln\xi\dot{)}u_a] +\dot{q}_a
+\dot{q}_fu^fu_a+(\ln\xi\dot{)}q_a\nonumber\\
&+&q^f\omega_{fa}+q^f\sigma_{fa}+\frac{1}{3}\theta q_a=\beta
q_a,\\
\dot{p}&+&\frac{2}{3}p\theta+\frac{2}{3}q^a\dot{u}_a
-\frac{2}{3}q^a(\ln\xi)_{,a}+\frac{2}{3}\pi^{ab}\sigma_{ba}=\beta
p,\\
2p\sigma_{ab}&+&2q_{(a}\dot{u}_{b)}-\frac{2}{3}q^f\dot{u}_fh_{ab}-
2q_{(a}(\ln\xi)_{,b)}-2u_{(a}q_{b)}(\ln\xi\dot{)}
+\nonumber\\\frac{2}{3}\theta\pi_{ab}
&+&\frac{2}{3}q^f(\ln\xi)_{,f}h_{ab}+\dot{\pi}_{ab}
+2\dot{\pi}_{c(a}u_{b)}u^c+2\pi_{f(a}\omega^f_{b)}\nonumber\\
&+&2\pi_{f(a}\sigma^f_{b)}
-\frac{2}{3}\pi^c_f\sigma^f_{c}h_{ab}=\beta \pi_{ab}.
\end{eqnarray}
\par\noindent
\textbf{Proof:}
\par\noindent
First we assume that timelike CMC exists in this spacetime and prove
that the conditions given by Eq.(4.1)-(4.4) are satisfied.

When we substitute value of the energy-momentum tensor for
anisotropic fluid from Eq.(2.19) in Eq.(2.9), we have
\begin{eqnarray}
(\rho&+&p\dot{)}u_a u_b+2(\rho+p) u_{(a} \dot{u}_{b)}+\dot{p}g_{ab}
+2q_{(a}\dot{u}_{b)}+2\dot{q}_{(a}u_{b)}
+\dot{\pi}_{ab}\nonumber\\
&-& 2\rho u_{(a}(ln\xi)_{,b)} -2q_{(a}(ln\xi)_{,b)}+2pu_{(a ;
b)}+2q_fu_{(a}u^f_{; b)}
+2\pi_{f(a}u^f_{;b)}\nonumber\\
&=&\beta[(\rho+p) u_a u_b +pg_{ab}+2q_{(a}u_{b)}+\pi_{ab}].\nonumber
\end{eqnarray}
This implies that
\begin{eqnarray}
[\dot{\rho}&+&2\rho(\ln\xi\dot{)}]u_au_b
+[2\rho(\dot{u}_c-(\ln\xi)_{,c})+2\dot{q}_c
+2(\ln\xi\dot{)}q_c\nonumber\\
&+&2q^f(\sigma_{fc}+\omega_{fc} +\frac{1}{3}\theta
h_{fc})]u_{(a}h^c_{b)}+\{\dot{p}h_{cd}+2p(\sigma_{cd}
+\frac{1}{3}h_{cd})\nonumber\\
&+&\dot{\pi}_{cd} +\frac{2}{3}\theta \pi_{cd}
+2q_{(c}[\dot{u}_{d)}-(\ln\xi)_{,d}] +2\pi_{f(c}[w^f_{d)}
+\sigma^f_{d)}]\}h^c_{a}h^d_{b}\nonumber\\
&=&\beta[\rho u_a u_b +ph_{ab}+2q_{(a}u_{b)}+\pi_{ab}].
\end{eqnarray}
Contracting Eq.(4.5) in turn with $u^au^b,~u^ah_c^b~h^{ab}$ and
$h^a_ch^b_d-\frac{1}{3} h^{ab}h_{cd}$, we obtain
\begin{eqnarray}
\dot{\rho}&+&2\rho(\ln\xi\dot{)}=\beta \rho,\\
\rho h^b_c&[&\dot{u}_b-(ln\xi)_{,b}] +\dot{q}_c
+(\ln\xi\dot{)}q_c+\dot{q}_fu^fu_c+q_f\omega^f_{c}\nonumber\\
&+&q_f\sigma^f_{c}+\frac{1}{3}\theta q_c=\beta q_c,\\
3\dot{p}&+&2p\theta+2\dot{u}_fq^f-2q^f(\ln\xi){,f}
+2\pi^{bf}\sigma_{fb}=3\beta p,\\
(h^a_ch^b_d&-&\frac{1}{3}h^{ab}h_{cd})
[2q_{(a}\dot{u}_{b)}+\dot{\pi}_{ab}
-2q_{(a}(\ln\xi)_{,b)}+2p\sigma_{ab}\nonumber\\
+2\pi_{f(a}\omega^f_{b)}&+&2\pi_{f(a}\sigma^f_{b)}+\frac{2}{3}\theta\pi_{ab}]
=\beta(h^a_ch^b_d-\frac{1}{3}h^{ab}h_{cd})\pi_{ab}.
\end{eqnarray}
Since
\begin{eqnarray*}
(h^a_ch^b_d-\frac{1}{3}h^{ab}h_{cd})\sigma_{ab}&=&
\sigma_{cd},\\
(h^a_ch^b_d-\frac{1}{3}h^{ab}h_{cd})(q_a\dot{u}_b+q_b\dot{u}_a)
&=&2q_{(c}\dot{u}_{d)}-\frac{2}{3}q^b\dot{u}_bh_{cd},\\
(h^a_ch^b_d-\frac{1}{3}h^{ab}h_{cd})\dot{\pi}_{ab}
&=&\dot{\pi}_{cd}+2\dot{\pi}_{b(c}u_{d)}u^b,\\
(h^a_ch^b_d-\frac{1}{3}h^{ab}h_{cd})
(\pi_{fa}\omega^f_{b}+\pi_{fb}\omega^f_{a})
&=&2\pi_{f(c}\omega^f_{d)},\\
(h^a_ch^b_d-\frac{1}{3}h^{ab}h_{cd})
(\pi_{fa}\sigma^f_{b}+\pi_{fb}\sigma^f_{a})
&=&2\pi_{f(c}\sigma^f_{d)}-\frac{2}{3}h_{cd}
\pi^{ab}\sigma_{ab},\\
(h^a_ch^b_d-\frac{1}{3}h^{ab}h_{cd})
(q_a(\ln\xi)_{,b}+q_b(\ln\xi)_{,a})
&=&2q_{(c}(\ln\xi)_{,d)}+2u_{(c}q_{d)}
(\ln\xi\dot{)}\nonumber\\
&-&\frac{2}{3}h_{cd}q^b(\ln\xi)_{,b},\\
(h^a_ch^b_d-\frac{1}{3}h^{ab}h_{cd})\pi_{ab}&=&\pi_{cd}.
\end{eqnarray*}
Thus Eq.(4.9) takes the following form
\begin{eqnarray}
2p\sigma_{ab}&+&2q_{(a}\dot{u}_{b)}-\frac{2}{3}q^f\dot{u}_fh_{ab}-
2q_{(a}(\ln\xi)_{,b)}-2u_{(a}q_{b)}(\ln\xi\dot{)}
+\frac{2}{3}\theta\pi_{ab}\nonumber\\
&+&\frac{2}{3}q^f(\ln\xi)_{,f}h_{ab}+\dot{\pi}_{ab}
+2\dot{\pi}_{c(a}u_{b)}u^c+2\pi_{f(a}\omega^f_{b)}
+2\pi_{f(a}\sigma^f_{b)}\nonumber\\
&-&\frac{2}{3}\pi^{cd}\sigma_{cd}h_{ab}=\beta \pi_{ab}.
\end{eqnarray}
\begin{description}
\item{$(i)$} Eq.(4.6) is the same as Eq.(4.1).
\item{$(ii)$} Eq.(4.2) is obtained by expanding Eq.(4.7).
\item{$(iii)$} Dividing Eq.(4.8) by 3, we get the condition (4.3).
\item{$(iv)$} Condition (4.4) is the same as Eq.(4.10).
\end{description}
Conversely, if the conditions (4.1)-(4.4) are satisfied, then there
must exist a timelike CMC in an anisotropic fluid spacetime. For
this purpose, we consider the left hand side of Eq.(4.5)
\begin{eqnarray*}
[\dot{\rho}&+&2\rho(\ln\xi\dot{)}]u_au_b
+[2\rho(\dot{u}_c-(\ln\xi)_{,c})+2\dot{q}_c
+2q^f(\sigma_{fc}+\omega_{fc}\nonumber\\
&+&\frac{1}{3}\theta h_{fc})]u_{(a}h^c_{b)}
+\{\dot{p}h_{cd}+2p(\sigma_{cd} +\frac{1}{3}h_{cd})+\dot{\pi}_{cd}
+\frac{2}{3}\theta \pi_{cd}\nonumber\\
&+&2q_{(c}[\dot{u}_{d)}-(\ln\xi)_{,d}]
+2\pi_{f(c}[w^f_{d)}+\sigma^f_{d)}]\}h^c_{a}h^d_{b}.
\end{eqnarray*}
Substituting the values from Eqs.(4.1)-(4.3) in the above
expression, we obtain
\begin{eqnarray}
2p\sigma_{ab}&+&2q_{(a}\dot{u}_{b)}-\frac{2}{3}q^f\dot{u}_fh_{ab}-
2q_{(a}(\ln\xi)_{,b)}-2u_{(a}q_{b)}(\ln\xi\dot{)}\nonumber\\
&+&\frac{2}{3}\theta\pi_{ab}
+\frac{2}{3}q^f(\ln\xi)_{,f}h_{ab}+\dot{\pi}_{ab}
+2\dot{\pi}_{c(a}u_{b)}u^c+2\pi_{f(a}\omega^f_{b)}\nonumber\\
&+&2\pi_{f(a}\sigma^f_{b)}
-\frac{2}{3}\pi^{cd}\sigma_{cd}h_{ab}+\beta[(\rho+p)u_au_b
+\frac{1}{2}(\rho-p+2\Lambda)g_{ab}\nonumber\\&+&2q_{(a}u_{b)}].
\end{eqnarray}
If we make use of Eq.(4.4), it follows that
$$\beta(\rho u_a u_b +ph_{ab}+2q_{(a}u_{b)}+\pi_{ab})$$
which is equal to the right hand side of Eq.(4.5). Hence the
conditions (4.1)-(4.5) are necessary and sufficient for anisotropic
fluid spacetime to admit a timelike CMC.

It is interesting to note that if we replace
$\frac{1}{2}(\rho+3p-2\Lambda)$ by $\rho$ and
$\frac{1}{2}(\rho-p+2\Lambda)$ by $p$ in Eqs. (3.1) to (3.4) then we
obtain Eqs.(4.1) to (4.4). This means that we can determine CMC from
CRC.\\
\par\noindent
\textbf{Proposition 4:}
\par\noindent
Anisotropic fluid with the energy-momentum tensor, given by
Eq.(2.19), admits a SpCMC $\xi^a=\xi x^a$ if and only if
\begin{eqnarray}
\rho^*&+&2\rho(\dot{u}^cx_c)-2(q^cx_c)(\ln\xi\dot{)}-2(q^cN_c)
+2q^cx_cu^{*d}x_d
=\beta \rho,\\
p[x^*_a &+&(\ln\xi)_{,a}-(\ln\xi)^*x_a] +(\pi_{ab}x^b)^*
-\pi^*_{cd}x^cx^dx_a
+\pi_{cd}x^cx^d(\ln\xi)_{,a}\nonumber\\&-&2\pi_{cd}x^cx^d(\ln\xi)^*x_a
+\pi_{ba}x^b(\ln\xi)^*+x^cq_c(\ln\xi)^*u_a-\pi_{cd}x^{*c}x^dx_a\nonumber\\
&-&\pi_{cd}\dot{x}^cx^du_a +x_c\dot{u}^bq_dx^bu_a
+\pi^{cd}x_dA_{ac}+q_cx^d\omega{cd}u_a\nonumber\\
&=&\beta[\pi_{ab}x^b-\pi_{cd}x^cx^dx_a
+q_bx^bu_a],\\
pN_a&+&2\rho\omega_{ab}x^b
+q_a\dot{x}^bu_b-q_bx^b\dot{x}^cu_cx_a-q^*_a+q^*_cx^cx_a
+q^*_bu^bu_a \nonumber\\&+&\pi^*_{ab}u^b+\pi_{cd}u^{*c}x^dx_a
-q_bx^b(\ln\xi)_{,a} -q_bx^b(\ln\xi\dot{)}u_a-q_bx^b(\ln\xi)^*x_a\nonumber\\
&+&\pi_{ab}x^b(\ln\xi\dot{)}-\pi_{cd}x^cx^d(\ln\xi\dot{)}x_a
+\pi_{ab}\dot{x}^b-\pi_{cd}x^c\dot{x}^dx_a-q^bA_{cb}\nonumber\\
&=&\beta(-q_a+q_bx^bx_a),\\
4p(\ln\xi)^*&-&2p\varepsilon+3\pi^*_{ab}x^ax^b+6\pi_{ab}x^bx^a(\ln\xi)^*
+2\pi_{ab}x^ax^{*b}\nonumber\\
&-&2x_a\pi^{ba}(\ln\xi)_{,b}-4x_aq^b\omega_{ab}-2\pi^{ab}A_{ab}
=3\beta\pi_{ab}x^ax^b,\\
2pS_{ab}&+&2\pi_{c(a}A^c_{b)}-2\pi^{cd}x_dA_{c(a}x_{b)}
+2\dot{x}^c\pi_{c(a}u_{b)}-2H_{ab}\pi^{cd}A_{cd}\nonumber\\
&+&\pi^*_{ab}+2u^c\pi^*_{c(a}u_{b)}
-2x^c\pi^*_{c(a}x_{b)}+\pi^*_{cd}x^cx^dx_ax_b
-2\pi^*_{cd}x^cu^dx_{(a}u_{b)}\nonumber\\
&+&\frac{1}{2}H_{ab}\pi^*_{cd}x^cx^d
-4x^cq_{(a}\omega_{b)c}+4x^cq_dx^dx_{(a}\omega_{b)c}
+2H_{ab}\omega_{cd}q^cx^d\nonumber\\&+&2x^c \pi_{c(a}(\ln\xi)_{,b)}
-2x^cx^d\pi_{cd}(\ln\xi)_{,(a}x_{b)}
+2(\ln\xi\dot{)}\pi_{c(a}u_{b)}x^c\nonumber\\
&-&2\pi_{cd}x^cx^d(\ln\xi\dot{)}x_{(a}u_{b)}
-2x^c(\ln\xi)^*\pi_{c(a}x_{b)}+2x^c\pi_{cd}x^d
(\ln\xi)^*x_ax_b\nonumber\\
&-&H_{ab}[x_c\pi^{cd}(\ln\xi)_{,d} -\pi_{cd}x^cx^d(\ln\xi)^*]
=\beta[\pi_{ab}-2x^c\pi_{c(a}x_{b)}\nonumber\\
&+&\pi_{cd}x^cx^dx_ax_b +\frac{1}{2}H_{ab}\pi_{cd}x^cx^d].
\end{eqnarray}
The proof follows similarly as for Proposition $3$.

\section{Outlook}

Physically, there is a close connection of inheriting CKVs with the
relativistic thermodynamics of fluids since for a distribution of
massless particles in equilibrium, the inverse temperature function
is inheriting CKV. This paper deals with the fundamental question of
determining when the symmetries of the geometry is inherited by all
the source terms of a prescribed matter tensor of EFEs. We have
investigated timelike and spacelike CRCs and CMCs of anisotropic
fluid using a particular procedure.

We formulate conditions defining the CRCs and CMCs for anisotropic
fluid. It is mentioned here that when we take $\pi_{ab}=0=q_a$ in
anisotropic fluid, the conditions for timelike and spacelike CRC
reduce to the conditions of perfect fluid [25]. Also, for
$\pi_{ab}=0,~q_a=0$, the conditions for timelike and spacelike CMC
reduce to the conditions of perfect fluid [23]. Further, for
$\alpha=0$, we obtain the conditions of RCs and MCs in each case of
spacelike and timelike CRCs and CMCs for the perfect fluid. This
shows that our results provide the generalization of the results
already available in the literature. It is worthwhile to note that
if we replace $\frac{1}{2}(\rho+3p-2\Lambda)$ by $\rho$ and
$\frac{1}{2}(\rho-p+2\Lambda)$ by $p$, then we can determine CMC
from CRC. We have obtained the conditions for the existence of CRCs
and CMCs in the models of anisotropic fluid. These conditions can be
used as restriction for the EFEs. Since the non-linearity of EFEs
ceases to extract their exact solution, the restricted equations may
give interesting solution in respective spacetimes. It can be shown
that a string fluid is the simplest example of anisotropic fluid
with vanishing heat flux. It would be interesting to extend this
procedure to the combination of two perfect fluids. This work is in
preparation [26] and will appear elsewhere.

\vspace{2cm}

{\bf References}

\begin{description}

\item{[1]} Oliver, D.T., and Davis, W.R.: Ann. Inst. Henri Poincare \textbf{3}(1977)399.

\item{[2]} Oliver, D.T., and Davis, W.R.: Gen. Rel. Grav. \textbf{8}(1979)905.

\item{[3]} Letelier, P.S.: Phys. Rev. \textbf{D22}(1980)807.

\item{[4]} Herrera, L. and Ponce de, L.J.: J. Math. Phys.
\textbf{26}(1985)778.
\item{[5]} Herrera, L. and Ponce de, L.J.: J. Math. Phys.
\textbf{26}(1985)2018.
\item{[6]} Herrera, L. and Ponce de, L.J.: J. Math. Phys. \textbf{26}(1985)2847.

\item{[7]} Maartens, R., Mason, D.P. and Tsamparlis, M.: J. Math. Phys.
\textbf{27}(1986)2987.

\item{[8]} Carot, J., Coley, A.A., and Sintes, A.: Gen. Rel. Grav.
\textbf{28}(1986)311.

\item{[9]} Coley, A.A. and Tupper, B.O.J.: J. Math. Phys. \textbf{30}(1989)2616.

\item{[10]} Duggal, K.L.: J. Math. Phys. \textbf{33}(1992)2989.

\item{[11]} Duggal, K.L.: Acta Appl. Math. \textbf{31}(1993)225.

\item{[12]} Greenberg, P.S.: J. Math. Anal. Appl. \textbf{30}(1970)128.

\item{[13]} Mason, D.P. and Tsamparlis, M.: J. Math. Phys.
\textbf{26}(1985)2881.

\item{[14]} Yavuz, I. and Yilmaz, I.: Gen. Rel. Grav.
\textbf{29}(1997)1295.

\item{[15]} Yilmaz, I., Tarhan, I., Yavuz, I.,
Baysal, H. and Camci, U.: Int. J. Mod. Phys. \textbf{D8}(1999)659.

\item{[16]} Yilmaz, I.: Int. J. Mod. Phys. \textbf{D10}(2001)681.

\item{[17]} Tsamparlis, M. and Mason, D.P.: J. Math. Phys. \textbf{31}(1990)1707.

\item{[18]} Tsamparlis, M.: J. Math. Phys. \textbf{33}(1992)1472.

\item{[19]} Baysal, H. and  Yilmaz, I.: Class Quantum Grav. \textbf{19}(2002)6435.

\item{[20]} Baysal, H. and Yilmaz, I.: Turk J. Phys. \textbf{27}(2003)83.

\item{[21]} Saridakis, E. and Tsamparlis, M.: J. Math. Phys. \textbf{32}(1991)1541.

\item{[22]} Tsamparlis, M.: Gen. Rel. Grav. {\bf 38}(2006)279.

\item{[23]} Sharif, M. and Sheikh, Umber: Int. J. Mod. Phys.
{\bf A21}(2006)3213.

\item{[24]} Stephani, Hans: {\it General Relativity: An Introduction to the Theory of
the Gravitational Field} (Cambridge University Press, 1990).

\item{[25]}  Duggal, K.L. and Sharma R.: {\it Symmetries of Spacetime
and Riemann Manifold} (Kluwer Academic Publisher, 1999).

\item{[26]} Sharif, M. and Tehseen, Naghmana: \emph{Conformal Ricci and Matter Collineations
for Two Perfect Fluids}, submitted for publication.
\end{description}

\end{document}